\documentclass[aps,pre,twocolumn,groupeaddress]{revtex4-1}
\usepackage[english]{babel}
\usepackage[utf8x]{inputenc}
\usepackage{amsmath,amsfonts,amssymb,amsbsy,eucal,dsfont,bbold}
\usepackage{natbib}
\usepackage{graphicx}

\begin{document}

\title{Granular beads in a vibrating, quasi two-dimensional cell:\\The true shape of the effective pair potential}

\author{Gustavo M. Rodr\'{i}guez-Li\~{n}\'{a}n}
\email[]{grodriguezlinan@gmail.com}
\affiliation{División de Ciencias e Ingenierías, Universidad de Guanajuato, 37150 León, Guanajuato, Mexico}

\author{Marco Heinen}
\email[]{mheinen@fisica.ugto.mx}
\affiliation{División de Ciencias e Ingenierías, Universidad de Guanajuato, 37150 León, Guanajuato, Mexico}

\date{\today}
\begin{abstract}
Steady-state pair correlations between inelastic granular beads in a vertically shaken,
quasi two-dimensional cell can be mapped onto the particle correlations in a truly two-dimensional reference fluid
in thermodynamic equilibrium.
Using Granular Dynamics simulations and Iterative Ornstein--Zernike Inversion,
we demonstrate that this mapping applies in a wide range of particle packing fractions and restitution coefficients,
and that the conservative reference particle interactions are simpler than it has been reported earlier.
The effective potential appears to be a smooth, concave function of the particle distance $r$.
At low packing fraction, the shape of the effective potential is compatible with a one-parametric
fit function proportional to $r^{-2}$.    
\end{abstract}

\maketitle
\section{Introduction}\label{sec:Intro}

Agitated granular materials tend to exhibit intricate phenomena such as pattern formation
\cite{Aranson2006}, collapse \cite{vanderMeer2002} or segregation \cite{Sanders2004, Rosato1987}.
In quasi two-dimensional systems it is not uncommon to observe two coexisting phases such as condensed clusters of particles
surrounded by a gas-like phase \cite{Olafsen1998, Roeller2011, Prevost2004, Risso2018}. For freely cooling
systems of inelastic particles studied \emph{in silico,} it has been reported that particles tend to
form clusters inside which the rate of energy dissipation exceeds that in the rest of the system,
in a process known as clustering instability \cite{Goldhirsch1993}.

Even though granular materials are systems far from equilibrium, several authors have proposed the
introduction of effective interactions among particles to describe the observed phase separation and
segregation \cite{Ciamarra2006, Bordallo-Favela2009}. Effective potentials have been
calculated for experimentally observed quasi two-dimensional systems of granular spheres under mechanical
agitation \cite{Bordallo-Favela2009} or under the effects of external, oscillating magnetic fields
\cite{Tapia-Ignacio2016, Donado2017}, by measuring the radial distribution function and inverting
it by means of the Percus--Yevick \mbox{(PY)} integral equation \cite{Percus1958}.
Following the same approach, Vel\'{a}zquez-P\'{e}rez and co-workers have studied the effect of the
interparticle coefficient of restitution on the shape of the effective potential, reporting
an increment of the effective particle attraction with decreasing values of the coefficient of restitution \cite{Velazquez-Perez2016}.
In their paper, they present a complicated shape of the attractive effective potential as a function of the 
particle separation distance.

In the present work we show that a simple inversion of the \mbox{PY} integral equation is insufficient
for obtaining the correct form of the effective potential in most granular systems. Instead, we propose
the use of the novel Iterative Ornstein--Zernike Inversion \mbox{(IO--ZI)} method \cite{Heinen2018} which is shown here to
yield more reliable and simpler forms of the effective potential.

This paper is organized as follows:
In Sec.~\ref{sec:Granul_Sim}, we describe our Granular Dynamics simulations.
The \mbox{IO--ZI} method for calculating the effective pair potential of a two-dimensional reference fluid is
explained in Sec.~\ref{sec:IO--ZI}, including a subsection~\ref{subsec:IO--ZI_validation} in which the
method is validated by test cases. Our results for the effective pair potential are reported in
Sec.~\ref{sec:Results}, which is followed by the conclusions.

\section{Granular Dynamics Simulations}\label{sec:Granul_Sim}

\begin{figure}
\centering
\includegraphics[width=.96\columnwidth]{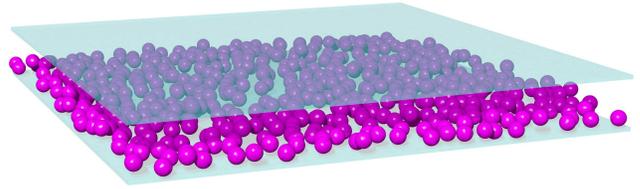}
\vspace{0em}
\caption{\label{fig:Snapshot} (Color online) A snapshot of our Granular \mbox{Dynamics} simulation for $\phi = 0.4$, $\epsilon = 0.7$.
The Cartesian box dimensions are $L \times L \times 3\sigma$ with $L / \sigma = \sqrt{\pi N / 4 \phi} ~ \approx 31.7$.}
\end{figure}

Figure~\ref{fig:Snapshot} features a representative snapshot from one of our Granular Dynamics simulations.
All simulations are for monodisperse systems of $N = 512$ spherical particles with diameter $\sigma$, confined
between two horizontal plates at $z = \delta z(x, y)$ and $z = 3\sigma + \delta z(x, y)$. Including the gentle sinusoidal
surface roughness $\delta z(x,y) = 10^{-3} \times \sigma \times \left[ \sin (\psi x) + \sin (\psi y) \right]$ with $\psi \sigma = 210$
on the plates helps to avoid a suppression of the $x$- and $y$-components of the spheres'
velocities due to friction between the particles and the plates \cite{Perera-Burgos2010}.
Our choice of the parameter $\psi$ corresponds to a surface roughness wavelength that is much shorter than $\sigma$,
resulting in quasi-random lateral velocity kicks.

Periodic boundary conditions are applied in the
Cartesian $x$- and $y$-directions, and the particles have three translational and three rotational
degrees of freedom. \mbox{Newton's} equation of motion is integrated in time by means of a Verlet
algorithm with a velocity-prediction step \cite{Perez2008}. Forces that act orthogonal to the
particle surfaces are modeled by a spring-dashpot model \cite{Shafer1996}, whereas tangential
interactions are modeled as Coulomb friction for the sake of simplicity in calculations.
The orthogonal forces are characterized by the
restitution coefficients $\epsilon$ and $\epsilon_w$ in case of particle-particle and particle-wall
collisions, respectively. In all our simulations, the particle-wall restitution coefficient
$\epsilon_w = 0.9$ is assumed. For the particle-particle normal restitution coefficient we have
used the three values $\epsilon = 0.5, 0.7$ and $0.9$. The tangential forces in particle pairs and
between particles and walls are both characterized by the tangential friction coefficient $\mu =
0.4$ in all our simulations.

In an initialization step, the particles are placed at random vertices of a horizontal,
two-dimensional triangular lattice with a lattice constant of $1.001\sigma$, at the center plane $z
= 3\sigma/2$ between the confining plates. All spheres are assigned random velocity vectors $\boldsymbol{v}_0$
with magnitudes in the range $0 < \left| \boldsymbol{v}_0 \right| < 8\times 10^{-5} ~\sigma/ \delta t$,
and random angular velocity vectors $\boldsymbol{\omega}_0$ with magnitudes in the range
$0 < \left| \boldsymbol{\omega}_0 \right| < \sqrt{3} \times 10^{-10} ~\text{rad} / \delta t$,
where $\delta t$ is the time step of the numerical integration scheme.
The confining plates are then moved sinusoidally in the $z$-direction with an amplitude $A =
0.012678~\sigma$ and a frequency $\nu = 1.4 \times 10^{-4} / \delta t$. The particles are affected
by a gravitational acceleration $g$ in the negative $z$-direction.
Setting the value of $g = 981~\text{cm} / \text{s}^{2}$, $\sigma =  0.5$~cm and $\delta t = 2\times 10^{-6}$~s, 
it is possible to express all simulation parameters in cgs units,
so that $\nu = 70$~Hz and $A = 0.006339$~cm.
Such parameters are realistic for experimental systems \cite{Bordallo-Favela2009}. The reduced,
dimensionless peak acceleration of the plates is $\Gamma = A(2\pi\nu)^2/g = 1.25$, and we define
a quasi-two-dimensional particle packing fraction as $\phi = (\pi N \sigma^2) / (4 L^2)$,
where $L$ is the simulation box length in the $x$- and $y$-directions.
We have performed simulations for packing fractions $\phi = 0.2, 0.4$ and $0.5$.

After a short initial transient, the simulations enter a steady state that appears stationary if short-time averages are considered.
In this steady state, the particles rebound vertically and acquire horizontal velocity components due to the surface
undulations of the confining plates and also via particle-particle collisions. A snapshot of all particle positions was stored
after every 16,667-th time step, corresponding to an interval of $1/30$~s between subsequent recordings.
A total number of 2,000 snapshots was recorded for each simulation, with an exception being the system at $\phi = 0.2$, $\epsilon = 0.5$
(lower right panel in Fig.~\ref{fig:Main_Figure} and Fig.~\ref{fig:rsquare}) for which we have recorded 10,000 snapshots.
From the snapshots we have calculated the projected two-dimensional radial distribution function
\begin{equation}\label{eq:gr_extraction}
g_{T}(r) = \dfrac{1}{N}
\left\langle
\sum\limits_{\substack{i,j = 1\\i \neq j}}^{N}
\delta \left( \boldsymbol{r}^{\parallel} - \boldsymbol{r}^{\parallel}_{i} + \boldsymbol{r}^{\parallel}_{j} \right)
\right\rangle 
\end{equation}
in terms of the Dirac $\delta$ distribution,
and the projected two-dimensional static (steady state) structure factor
\begin{equation}\label{eq:Sq_extraction}
S_{T}(q) = \dfrac{1}{N}
\left\langle
{\left[ \sum\limits_{i = 1}^{N} \cos ( \boldsymbol{q}^{\parallel} \cdot \boldsymbol{r}^{\parallel}_{i} )  \right]}^2 \hspace{-.4em} + 
{\left[ \sum\limits_{i = 1}^{N} \sin ( \boldsymbol{q}^{\parallel} \cdot \boldsymbol{r}^{\parallel}_{i} )  \right]}^2
\right\rangle
\end{equation}
where $\left\langle \ldots \right\rangle$ stands for the average over all snapshots,
$\boldsymbol{r}^{\parallel}_{i} = \left( \mathbb{1} - \hat{\boldsymbol{e}}_{z} \hat{\boldsymbol{e}}_{z} \right) \cdot \boldsymbol{r}_{i}$ 
is the projection of the position vector $\boldsymbol{r}_{i}$ of particle $i$ into the $(x, y)$-plane, and 
$\boldsymbol{q}^{\parallel} = \left( \mathbb{1} - \hat{\boldsymbol{e}}_{z} \hat{\boldsymbol{e}}_{z} \right) \cdot \boldsymbol{q}$ 
is the corresponding projection of the wave vector $\boldsymbol{q}$. The arguments $r = \left| \boldsymbol{r}^{\parallel} \right|$ and
$q = \left| \boldsymbol{q}^{\parallel} \right|$ of the correlation functions are the norms of the projected distance and wave vectors.
We have checked that all simulated systems are homogeneous and isotropic on average. 
The lower index '$T$' on both functions $g_{T}(r)$ and $S_{T}(q)$ stands for 'Target',
as we have used these functions as the target functions for the Iterative Ornstein--Zernike Inversion method,
described in Sec.~\ref{sec:IO--ZI}.

\section{Iterative Ornstein--Zernike Inversion}\label{sec:IO--ZI}
Iterative Ornstein--Zernike Inversion \mbox{(IO--ZI)} is a recently introduced inverse Monte Carlo method that allows to determine the reduced,
dimensionless pair potential $\beta u(r)$ of particles in thermodynamic equilibrium from their radial distribution function $g(r)$ and the static structure factor $S(q)$.
Here, $\beta = 1 / (k_B T)$ is the inverse thermal energy in terms of the Boltzmann constant $k_B$ and the absolute temperature $T$.
The interested reader is referred to Ref.~\cite{Heinen2018} for a comprehensive description of the IO--ZI method and its
validation for three-dimensional fluid systems. For brevity's sake, we explain here only the essential working principle of \mbox{IO--ZI}, and we mention
the differences between the algorithm in Ref.~\cite{Heinen2018} and the version for two-dimensional systems that we have used for the present work:

The \mbox{IO--ZI} method shares its underlying principle with the well-established, but less accurate Iterative Boltzmann Inversion \mbox{(IBI)} method \cite{Reith2003}.
In an initial step, a first estimate $\beta u_1(r)$ of the true potential $\beta u(r)$ is calculated via approximate, numerical inversion of the target correlation
functions $g_{T}(r)$ and $S_{T}(q)$ at known particle number density $n$. The reduced potential $\beta u_1(r)$ is then used in a strictly two-dimensional $(N,V,T)$
Metropolis Monte Carlo \mbox{(MC)} simulation from which the correlation functions $g_1(r)$ and $S_1(q)$ are extracted. The differences between $g_{T}(r)$ and $g_1(r)$ and
between $S_{T}(q)$ and $S_1(q)$ are the inputs for an iteration update rule by which the function $\beta u_1(r)$ is transformed into the next estimate $\beta u_2(r)$.
The latter serves as the reduced pair potential in a second \mbox{MC} simulation, resulting in $g_2(r)$ and $S_2(q)$. This sequence of potential adjustments and \mbox{MC}
simulations is continued until $g_n(r)$ and $S_n(q)$ are indistinguishable from $g_{T}(r)$ and $S_{T}(q)$, within the level of the stochastic noise floor.
At this point, $\beta u_n(r)$ constitutes the output of the \mbox{IO--ZI} (or the \mbox{IBI}) method.
 
Both the initial seed $\beta u_1(r)$ and the iteration update rule in \mbox{IO--ZI} rely on an approximation of the unknown bridge function \cite{Hansen_McDonald1986}
in the Ornstein--Zernike integral equation formalism. Different bridge function approximations, also known as closure relations, constitute different flavors of \mbox{IO--ZI}
such as Iterative Hypernetted Chain Inversion \mbox{(IHNCI)} which is based on the \mbox{HNC} closure \cite{Morita1958} or Iterative Percus-Yevick Inversion \mbox{(IPYI)},
based on the \mbox{PY} closure \cite{Percus1958}. The \mbox{IHNCI} algorithm has been published in Ref.~\cite{Heinen2018}, and the \mbox{IPYI} algorithm is obtained
if Eqs.~(8) and (9) from Ref.~\cite{Heinen2018} are replaced by the equations
\begin{equation}
\beta u_1(x) = \ln\left[g_{T}(x) - c_{T}(x)\right] - \ln\left[g_{T}(x)\right] \nonumber
\end{equation}
and 
\begin{equation}
\beta \mu_i(x) = \beta u_i(x) + \ln\left[ \dfrac{g_{T}(x) - c_{T}(x)}{g_{i}(x) - c_{i}(x)} \right] + \ln\left[ \dfrac{g_{i}(x)}{g_{T}(x)} \right],  \nonumber
\end{equation}
respectively.
Here, $x = r n^{1/d}$ is the dimensionless particle center-to-center distance in terms of the mean geometric particle distance $n^{-1/d}$.
The symbols $\mu_i(x)$ and $c_{T}(x)$ denote the output of a single Picard iteration of the \mbox{IPYI} algorithm and the target direct correlation function,
respectively. The meaning of both these quantities is discussed in great detail in Ref.~\cite{Heinen2018} and will not be repeated here for the sake of brevity.

The \mbox{IHNCI} and \mbox{IPYI} methods are surpassing the \mbox{IBI} method in terms of accuracy of the converged solution for the particle pair potential because
the initial seed and the iteration update rule in \mbox{IBI} are both based on the comparatively inaccurate approximation of the true pair potential by the potential
of mean force \cite{Hansen_McDonald1986}. Moreover, the \mbox{IO--ZI} methods make use of the information contained in the Fourier-space functions
$S_{T}(q)$ and $S_{i}(q)$ as well as the real space functions $g_{T}(r)$ and $g_{i}(r)$, whereas the \mbox{IBI} method relies on the real space information
from the radial distribution functions only.

The initial seed $\beta u_1(r)$ in \mbox{IHNCI} and \mbox{IPYI} is obtained via inversion of the \mbox{HNC} and \mbox{PY} integral equations, respectively.
We will therefore use the notation HNC Inversion \mbox{(HNCI)} and PY Inversion \mbox{(PYI)} for the numerical schemes that are obtained when only the 
initialization steps of \mbox{IHNCI} or \mbox{IPYI} are executed, and the subsequent MC simulations and iterative potential corrections are omitted.
The so-obtained \mbox{PYI} method has already been used \cite{Bordallo-Favela2009, Velazquez-Perez2016} to calculate effective potentials of granular beads in vibrated quasi-two-dimensional
cells, but we are going to demonstrate in Sec.~\ref{sec:Results} that the results from \mbox{PYI} and \mbox{HNCI} are not reliable as they contain a large
systematic error. Effective potentials of granular beads that have so far been published must therefore be challenged and re-checked in every particular case. 

As an additional technical comment, we note that the necessary inverse Fourier (or Hankel) transform $\mathcal{F}^{-1}$
of the isotropic direct correlation function $\tilde{c}(q)$ from wavenumber space into the real-space
function $c(r)$ should preferentially be carried out via the equation
\begin{equation}
c(x) = g(x) - 1 - \mathcal{F}^{- 1} \left\lbrace \dfrac{{\left[S(y) - 1\right]}^2}{S(y)} \right\rbrace (x) \nonumber
\end{equation}
\cite{Heinen2018, Heinen2011},
in which $y = q n^{-1/d}$ is a dimensionless wavenumber, and where
the Fourier integrand ${\left[S(y) - 1\right]}^2 / S(y)$ decays considerably
quicker as a function of $y$ than the integrand $\tilde{c}(y)$ in
$c(x) = \mathcal{F}^{-1} \left\lbrace \tilde{c}(y) \right\rbrace (x)$.
A fast decay of the Fourier integrand is a desirable feature as the correlation functions are typically only known in very limited ranges of the variables
$x$ and $y$. The Fourier transform is most accurately and conveniently carried out in arbitrary dimension by virtue of Hamilton's FFTLog algorithm
\cite{Hamilton2000, Hamilton_website}, which is based on Talman's original publication \cite{Talman1978}. 

All \mbox{IHNCI} and \mbox{IPYI} runs reported here were carried out with the generalized accelerated fixed-point iteration method originally proposed by Ng \cite{Ng1974, Heinen2014, Heinen2018}.
The \mbox{MC} simulations were performed on a graphics processing unit with ensemble averaging over $256$ statistically independent systems, each containing $256$ particles.
While this may appear to be a dangerously small particle number, our results confirm that it is large enough to avoid significant finite size effects on $\beta u(r), g(r)$ and $S(q)$.
In the validation and results sections \ref{subsec:IO--ZI_validation} and \ref{sec:Results} we will observe that the functions $g(r)$ and $S(q)$ from our 'forward direction' MC simulations
for twice the number of ($N =  512$) particles are perfectly reproduced in the inverse MC runs with $N =  256$.
The physical reason is that the particle interactions are short ranged.
Each one of the \mbox{IHNCI} and \mbox{IPYI} runs reported in subsection~\ref{subsec:IO--ZI_validation} and in Sec.~\ref{sec:Results} took $\sim 2$ hours to complete.
A \mbox{HNCI} or \mbox{PYI} run requires less than a second of runtime.

\subsection{Validation of \mbox{IO--ZI} for two-dimensional systems}\label{subsec:IO--ZI_validation}

\begin{figure}
\centering
\includegraphics[width=.85\columnwidth]{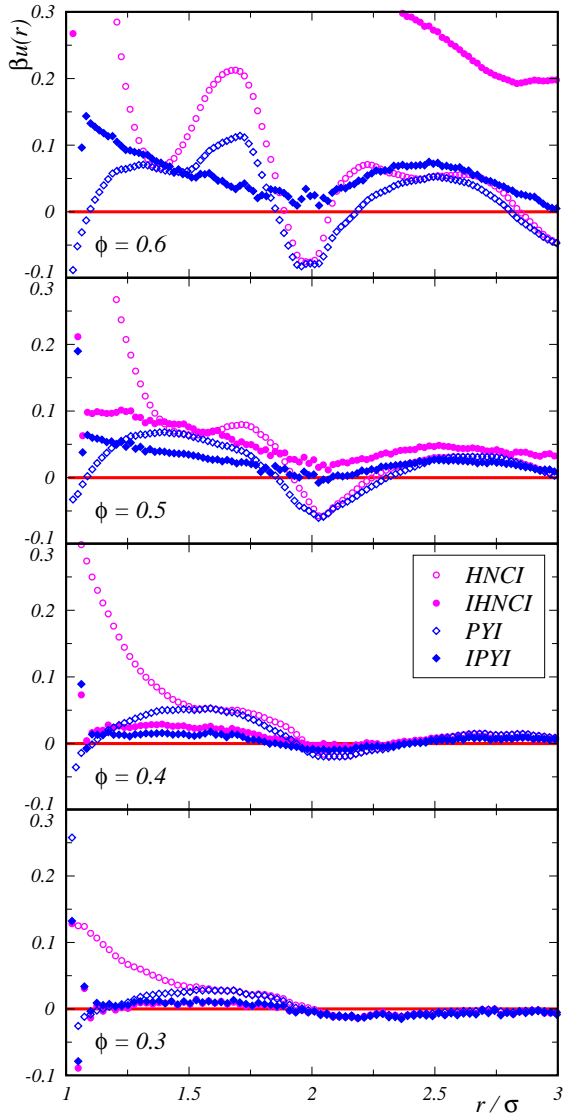}
\vspace{0em}
\caption{\label{fig:2D_HD_Tests}
(Color online) The \mbox{HNCI} (open pink circles), \mbox{IHNCI} (filled pink circles), \mbox{PYI} (open blue diamonds) and \mbox{IPYI} (filled blue diamonds)
methods are tested in their capabilities to reproduce the pair potential of hard disks in two dimensions (red horizontal lines),
at packing fractions $\phi = 0.3, 0.4, 0.5$ and $0.6$ (bottom panel to top panel).   
}
\end{figure}

Comprehensive validation tests of the \mbox{IO--ZI} method in its \mbox{IHNCI} flavor have been reported in Ref.~\cite{Heinen2018} for systems with
various types of particle pair potentials, but in three spatial dimensions only. Before applying \mbox{IHNCI} and \mbox{IPYI} in Sec.~\ref{sec:Results},
we validate both methods for the case of two-dimensional systems in the present subsection.  

Figure~\ref{fig:2D_HD_Tests} features the results from the \mbox{HNCI}, \mbox{IHNCI}, \mbox{PYI} and \mbox{IPYI} methods for four test cases in which the
target functions $g_{T}(r)$ and $S_{T}(q)$ are those of non-overlapping hard disks in two dimensions, at packing fractions $\phi = 0.3, 0.4, 0.5$ and $0.6$.
The functions $g_{T}(r)$ and $S_{T}(q)$ were calculated via Eqs.~\eqref{eq:gr_extraction} and \eqref{eq:Sq_extraction} in MC simulations of $512$ disks
with diameter $\sigma$, in two-dimensional square simulation boxes with periodic boundary conditions in both Cartesian directions.
The interaction potential $u(r > \sigma) = 0$ is represented by the horizontal red lines in Fig.~\ref{fig:2D_HD_Tests}. Any deviation from these lines
quantifies an inaccuracy of the \mbox{HNCI}, \mbox{IHNCI}, \mbox{PYI} or \mbox{IPYI} method. Note that \mbox{IHNCI} and \mbox{IPYI} are considerably more
accurate than \mbox{HNCI} and \mbox{PYI} in all studied cases, with the exception of the densest system at $\phi = 0.6$, where \mbox{IHNCI} fails 
dramatically. For all other systems at packing fractions $\phi = 0.5$ or less, the error of the converged reduced potentials $\beta u(r)$ from
\mbox{IHNCI} and \mbox{IPYI} stays below, or well below $0.1$ for practically all particle distances $r$. As one should expect, the \mbox{IPYI} method
is more accurate than the \mbox{IHNCI} method (and, likewise, \mbox{PYI} is more accurate than \mbox{HNCI}) in the hard disk test cases. This is due to the
well-known fact that the \mbox{PY} closure is more accurate for hard disks than the \mbox{HNC} closure \cite{Hansen_McDonald1986}. 

\begin{figure}
\centering
\vspace{1em}
\includegraphics[width=.85\columnwidth]{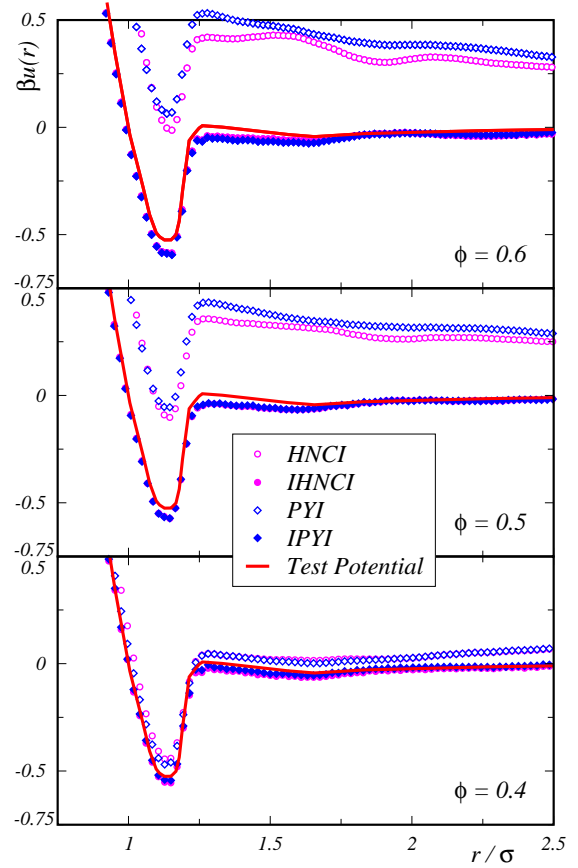}
\vspace{0em}
\caption{\label{fig:2D_Freestyle_Tests}
(Color online) Same as Fig.~\ref{fig:2D_HD_Tests}, but for a generic freehand-curve test potential (red curves), 
and at the packing fractions $\phi = 0.4, 0.5$ and $0.6$ (bottom panel to top panel).   
}
\end{figure}

For different interaction potentials, it is in general not known \textit{a priori} which one of the two closures -- HNC or PY -- is more accurate.
We have therefore conducted a set of three additional validation tests of \mbox{IHNCI} and \mbox{IPYI} with two-dimensional systems at packing fractions
$\phi = 0.4, 0.5$ and $0.6$, where the potential to be reproduced was taken from a digitalized free-hand curve that features strong repulsion at
distances $r < \sigma$, an attractive region of maximum depth $-0.5 k_B T$ in the region $\sigma < r < 1.25 \sigma$, and a quickly decaying, slightly
repulsive part at $r > 1.25 \sigma$. The results of these tests are shown in Fig.~\ref{fig:2D_Freestyle_Tests}, where the red solid curves represent the test
potential. The target functions $g_{T}(r)$ and $S_{T}(q)$ for \mbox{HNCI}, \mbox{IHNCI}, \mbox{PYI} and \mbox{IPYI} were extracted from MC simulations
of $512$ particles in two-dimensional square simulation boxes with periodic boundary conditions in both Cartesian directions, and with interactions described
by the test potential. As a result, we note that \mbox{IHNCI} and \mbox{IPYI} are considerably more accurate in reproducing the test potential than \mbox{HNCI} and \mbox{PYI},
especially at the two higher packing fractions $\phi = 0.5$ and $\phi = 0.6$. 

\begin{figure*}
\centering
\includegraphics[width=0.92\textwidth]{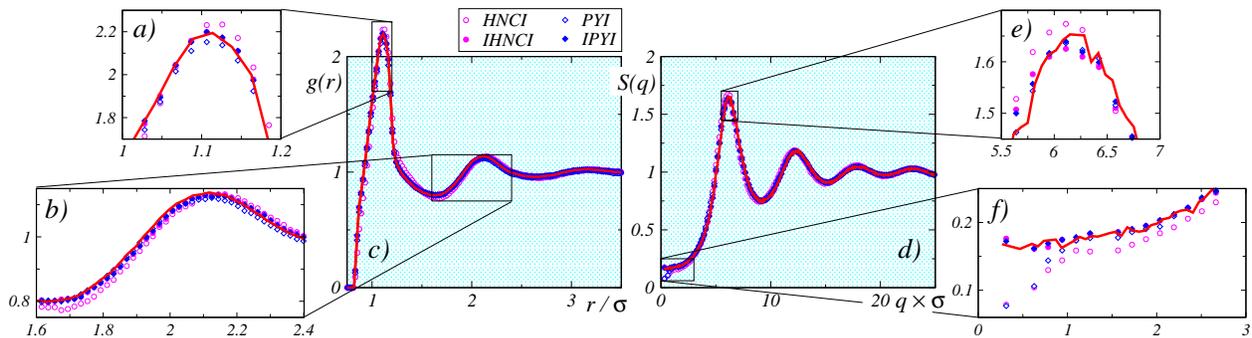}
\vspace{0em}
\caption{\label{fig:2D_Freestyle_gr_Sq}
(Color online) Radial distribution functions $g(r)$ and static structure factors $S(q)$ of the systems at packing fraction $\phi = 0.5$, and 
with reduced potentials plotted in the central panel of Fig.~\ref{fig:2D_Freestyle_Tests}.
Red solid curves represent the target correlation functions $g_{T}(r)$ and $S_{T}(q)$.}
\end{figure*}

The level of accuracy at which the target correlation functions $g_{T}(r)$ and $S_{T}(q)$ are reproduced by the
\mbox{HNCI}, \mbox{IHNCI}, \mbox{PYI} and \mbox{IPYI} methods is demonstrated in Fig.~\ref{fig:2D_Freestyle_gr_Sq}, which features our results
for the systems with reduced potentials plotted in the central panel of Fig.~\ref{fig:2D_Freestyle_Tests}:
All four inversion methods result in correlation functions $g(r)$ and $S(q)$ that are nearly identical to $g_{T}(r)$ and $S_{T}(q)$,
to a level at which the functions are almost indistinguishable within the stochastic noise floor of the simulation results.
Nevertheless, close observation of the correlation functions (as in panels a, b, e and f of  Fig.~\ref{fig:2D_Freestyle_gr_Sq})
reveals that \mbox{IHNCI} is ever so slightly more accurate in reproducing $g_{T}(r)$, $S_{T}(q)$ than \mbox{HNCI} is, 
and the same can be said about \mbox{IPYI} and its relation to \mbox{PYI}.
The minuscule differences between the correlation functions from \mbox{IHNCI} and \mbox{HNCI}, or between \mbox{IPYI} and \mbox{PYI},
are crucial, as they translate into stark differences between the reduced potentials.
This is a manifestation of the low practical usefulness of Henderson's theorem \cite{Henderson1974} as discussed in Refs.~\cite{Potestio2014, Heinen2018}:
In equilibrium fluids with pairwise additive particle interactions a bijective functional mapping $\beta u(r) \leftrightarrow \left[ g(r), S(q) \right]$
is guaranteed to exist, but the mapping is highly nonli	near in general. Large differences in $\beta u(r)$ may correspond to tiny differences
in $g(r)$ and $S(q)$ which complicates severely the calculation of $\beta u(r)$ from the correlation functions if these are only known within a
statistical error margin. This explains the severe failure of simple methods such as \mbox{HNCI} or \mbox{PYI}. More sophisticated methods 
such as \mbox{IHNCI}, \mbox{IPYI}, or alternative approaches such as pressure-corrected \mbox{IBI} \cite{Reith2003, Potestio2014} or
multistate \mbox{IBI} \cite{Moore2014} are required instead.

A few important characteristics of \mbox{IHNCI} and \mbox{IPYI} can be observed in both Figs.~\ref{fig:2D_HD_Tests} and \ref{fig:2D_Freestyle_Tests}:
Both methods are very accurate at small packing fractions and they gradually loose accuracy when the packing fraction is increased.
The packing fraction at which any one of the two methods starts to fail gravely can be estimated by comparison with the respective other method.
In other words, for cases where \mbox{IHNCI} and \mbox{IPYI} predict similar results, we have strong empirical evidence for the accuracy of both methods.
In converse cases where the results of \mbox{IHNCI} and \mbox{IPYI} differ markedly, neither of the two methods can be trusted.
We make use of the reassuring comparison between \mbox{IHNCI} and \mbox{IPYI} throughout the results section~\ref{sec:Results},
where the effective potentials for granular particles are calculated by both methods in all cases.

\section{Results}\label{sec:Results}

\begin{figure*}
\centering
\includegraphics[width=.9\textwidth]{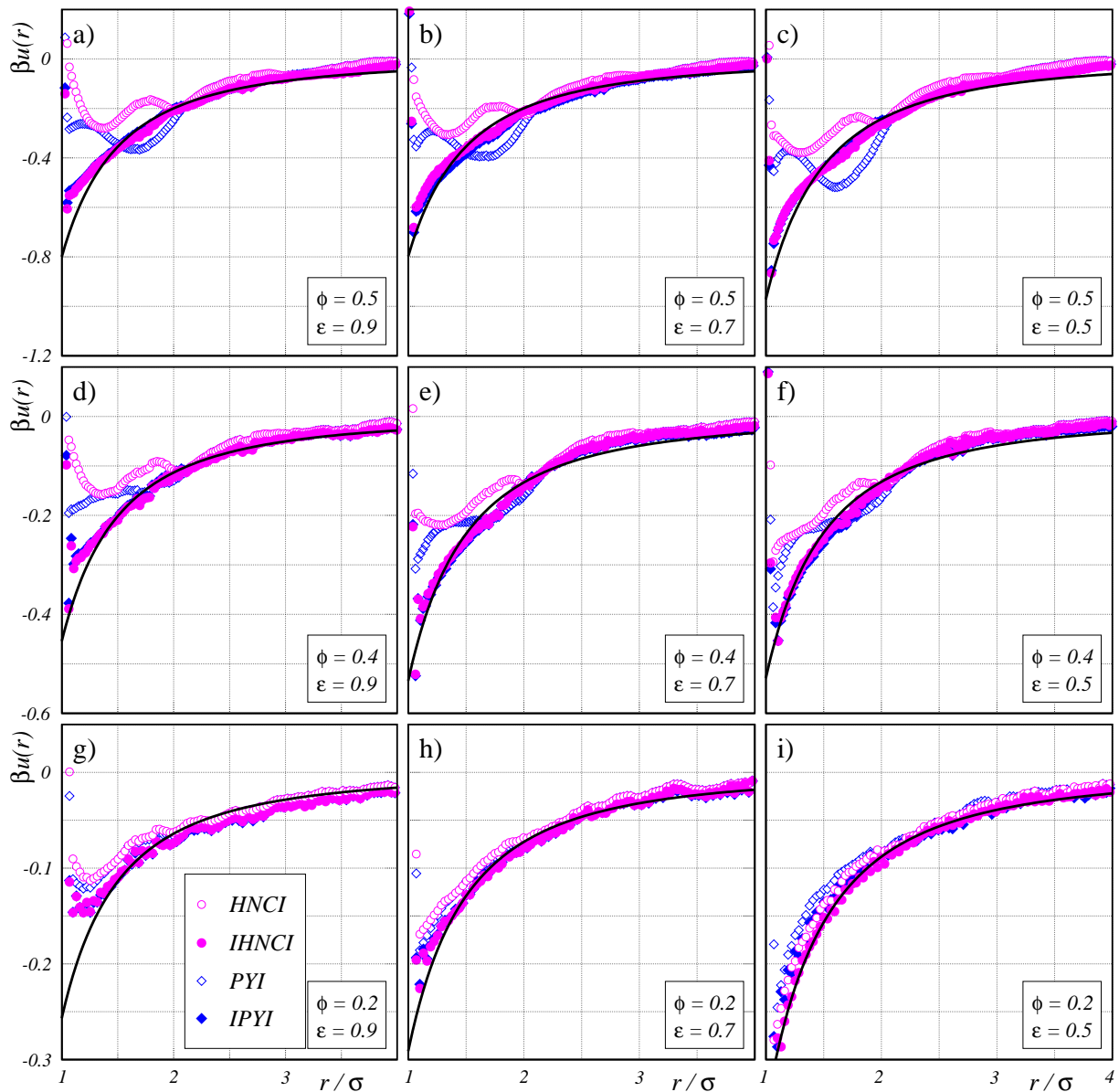}
\vspace{0em}
\caption{\label{fig:Main_Figure}
(Color online) 
Effective potentials for inelastic granular beads at packing fractions $\phi = 0.2$ (bottom row of panels), $\phi = 0.4$ (central row of panels) and $\phi = 0.5$ (top row of panels),
and for restitution coefficients $\epsilon = 0.9$ (left column of panels), $\epsilon = 0.7$ (central column of panels) and $\epsilon = 0.5$ (right column of panels).
Our results from the HNCI (open pink circles), IHNCI (filled pink circles), PYI (open blue diamonds) and IPYI (filled blue diamonds) methods are shown.
One-parametric functions of the form $\alpha / r^2$ (black curves) have been fitted to the IHNCI results in the range $1.25 < r / \sigma < 4$. 
The horizontal axis range is $1 < r/\sigma < 4$ is every panel, and the vertical axis range is varying by factors of $2$ from the bottom row to the center row,
and from the center row to the top row of panels.
}
\end{figure*}

Figure~\ref{fig:Main_Figure} features the main results of the present paper.
The \mbox{HNCI}, \mbox{IHNCI}, \mbox{PYI} and \mbox{IPYI} results for the reduced potentials $\beta u(r)$ in nine two-dimensional
equilibrium systems are plotted. The input (or target) functions $g_{T}(r)$ and $S_{T}(q)$ for the four inversion methods are those
that were obtained from our Granular Dynamics simulations as described in Sec.~\ref{sec:Granul_Sim}, for the restitution coefficients
$\epsilon = 0.9, 0.7$ and $0.5$ and the packing fractions $\phi = 0.2, 0.4$ and $0.5$. We have also conducted Granular Dynamics simulations
at $\phi = 0.6$ but we refrain from showing the results for the effective potentials here, as each one of the four inversion methods
is clearly failing at $\phi = 0.6$. Our observations in Fig.~\ref{fig:Main_Figure} are the following:

The \mbox{HNCI} and \mbox{PYI} results are in strong disagreement with each other and with the \mbox{IHNCI} and \mbox{IPYI} results,
for all but the most dilute systems at $\phi = 0.2$ (panels g, h and i of Fig.~\ref{fig:Main_Figure}). Both \mbox{HNCI} and \mbox{PYI}
are thus unreliable and should never be used in the determination of effective interaction potentials. 

Our \mbox{PYI} results for $\beta u(r)$ at $\phi = 0.4$ (panels d, e and f of Fig.~\ref{fig:Main_Figure})
resemble those in Fig.~4 of Ref.~\cite{Velazquez-Perez2016} as far as the shape of the functions is concerned, but the reduced
potentials in Ref.~\cite{Velazquez-Perez2016} are more strongly attractive, with a minimum value around $-1$ to $-1.5$, which is
approximately three times deeper than the minima of our results for $\beta u(r)$ at $\phi = 0.4$.
We presume that the reason for this quantitative disagreement might be a difference between the particle-wall restitution coefficients
$\epsilon_{w}$ of our Granular Dynamics simulations and those that were used in Ref.~\cite{Velazquez-Perez2016}. If the value
of $\epsilon_{w}$ was chosen smaller than our value of $\epsilon_{w} = 0.9$, then the effective, kinetic temperature of the Granular beads
in Ref.~\cite{Velazquez-Perez2016} can be expected to be lower than in our case, which would be in line with a larger value of $\beta$.
Unfortunately we are not in the position to test our presumption as the value of $\epsilon_{w}$ has not been reported in Ref.~\cite{Velazquez-Perez2016}.

Our \mbox{IHNCI} and \mbox{IPYI} results are in close agreement with each other, in all nine cases shown in Fig.~\ref{fig:Main_Figure},
which serves as a reassurance for the fidelity of both methods. A non-trivial finding is that both \mbox{IHNCI} and \mbox{IPYI} are
converging in all nine cases, and that the target correlation functions $g_{T}(r)$ and $S_{T}(q)$ of the out-of-equilibrium granular systems
are reproduced by the two methods (as we have checked in every case). This implies that there is indeed an equilibrium system with correlation
functions identical to those of the granular system in the entire parameter range $0.2 \leq \phi \leq 0.5$ and $0.5 \leq \epsilon \leq 0.9$.

The effective potentials from \mbox{IHNCI} and \mbox{IPYI} are attractive and follow a simple, monotonically increasing and concave shape in all cases,
with the exception of the system at $\phi = 0.2$ and $\epsilon = 0.9$ in panel g of Fig.~\ref{fig:Main_Figure}, where a gentle upturn of the reduced
potentials is observed at very close particle proximity. We cannot be sure about the statistical significance of that upturn and refrain
from over-interpreting it as a physical effect as it may just as well be a numerical artifact.
That the effective interactions are attractive is physically quite intuitive:
In a steady state with vanishing average particle currents, the normal velocity restitution causes an increase in particle number density
around any tagged particle, as it is also caused by attractive interactions in the effective equilibrium system with the same particle correlation functions.

\begin{figure}
\centering
\includegraphics[width=.88\columnwidth]{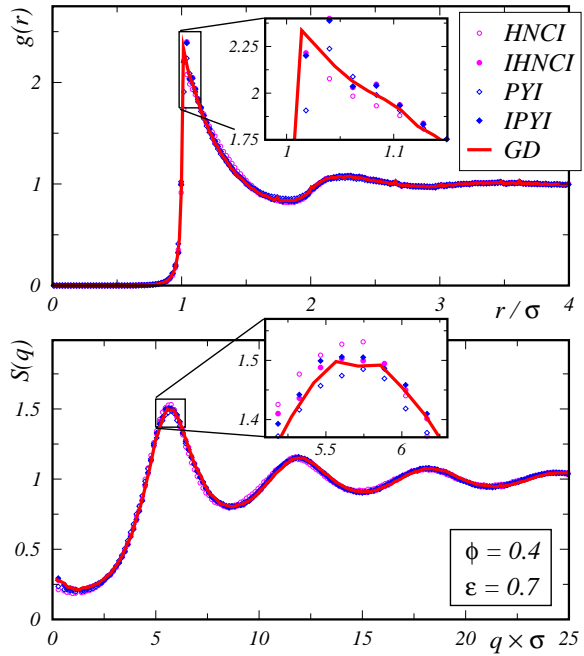}
\vspace{0em}
\caption{\label{fig:gr_Sq_granular}
(Color online) Radial distribution function (upper panel) and structure factor (lower panel) for $\phi = 0.4$ and $\epsilon = 0.7$.
Solid red curves are our Granular Dynamics \mbox{(GD)} results, extracted from the simulations via Eqs.~\eqref{eq:gr_extraction} and \eqref{eq:Sq_extraction}.
The corresponding effective potentials are plotted in the central panel of Fig.~\ref{fig:Main_Figure}.   
}
\end{figure}

The correlation functions for the system at $\phi = 0.4$, $\epsilon = 0.7$ (central panel `e' in Fig.~\ref{fig:Main_Figure}, also featured in Fig.~\ref{fig:Snapshot})
are shown in Fig.~\ref{fig:gr_Sq_granular}, where an upturn of $S(q)$ at small values of $q$ supports our finding of attractive effective interactions,
and $g(r < \sigma) \ll 1$ signals that the granular system is indeed nearly perfectly two-dimensional.

\begin{figure}
\centering
\vspace{2em}
\includegraphics[width=.82\columnwidth]{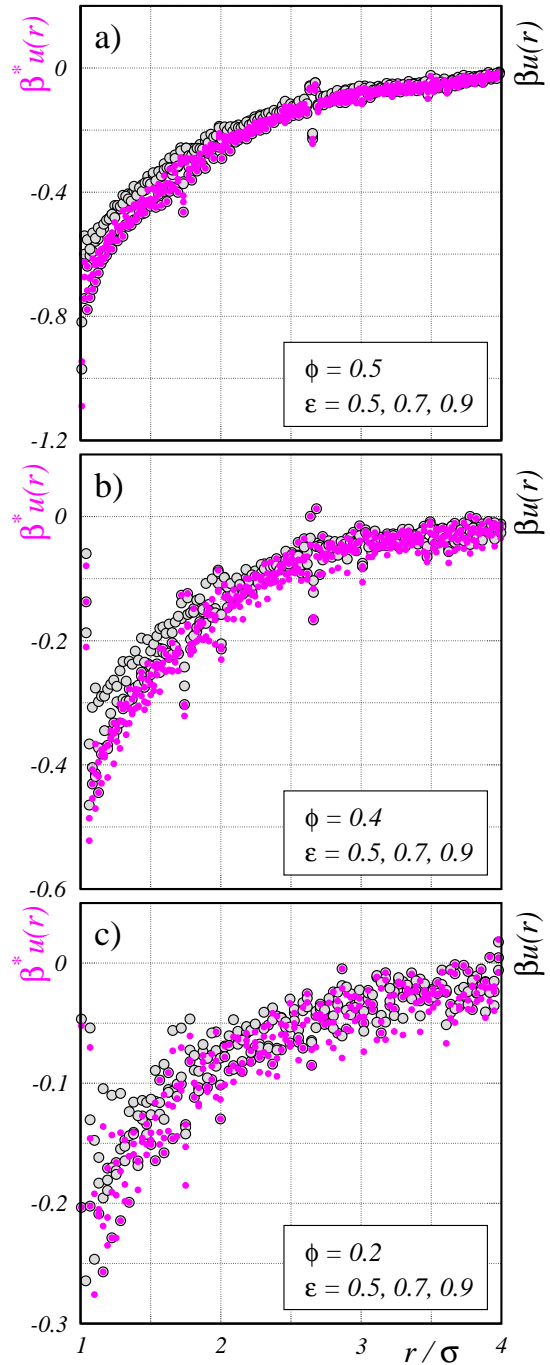}
\vspace{0em}
\caption{\label{fig:Tgranular_scaling} (Color online)
Black circles filled in gray: Converged \mbox{IHNCI} potentials from Fig.~\ref{fig:Tgranular_scaling}.
Pink circles: The same data, rescaled with respect to the effective granular temperature and the effective inverse thermal energy $\beta^*$, as described in the main text.
Every panel is for one packing fraction $\phi$. The results for different restitution coefficients $\epsilon$, corresponding to different values of $\beta^*$, are
overlaid in the panels. The vertical spread among the data narrows upon effective temperature rescaling.}
\end{figure}

Figure~\ref{fig:Tgranular_scaling} repeats all the converged \mbox{IHNCI} potentials from Fig.~\ref{fig:Main_Figure} as black circles filled in gray.
Every panel of Fig.~\ref{fig:Tgranular_scaling} is for one of the three packing fractions $\phi = 0.2, 0.4$ and $0.5$, as indicated in the panels a -- c.
The panels contain the results for three different restitution coefficients $\epsilon = 0.5, 0.7$ and $0.9$ in an overlaid manner, such that the spread among
the symbols of equal type indicates the difference between the results for equal packing fraction and for varying coefficient of restitution.
While the data for different $\epsilon$ appear to follow the same functional form (within statistical scatter), the observed vertical spread among the black/gray
circles indicates that different values of $\epsilon$ correspond to different values of the effective inverse thermal energy $\beta$. Keeping in mind that all our data
are for equal intensities of vertical shaking, this apparent spread in effective granular temperature is in line with the intuitive picture in which different
restitution coefficients $\epsilon$ correspond to different amounts of kinetic energy dissipation in the steady state.

As wee have checked, the distributions of the Cartesian velocity components parallel to the confining plates are nearly Maxwellian, with slight deviations from the
Maxwellian form for very slow and very fast velocities. This is in line with the observations that have been reported in several instances in the literature \cite{Olafsen1999, VanZon2004}.

An effective granular temperature was determined for each of the cases displayed in Figs.~\ref{fig:Main_Figure},~\ref{fig:Tgranular_scaling}, by 
fitting the Cartesian velocity histograms from our Granular Dynamics simulations to Maxwellian (Gaussian) functions,
using the variance $\sigma(\phi, \epsilon)$ of the distribution for each given pair of values $(\phi, \epsilon)$ as a fit parameter.
Assuming that the so-determined velocity variance is proportional to an effective granular Temperature, we have rescaled the data
with prefactors \mbox{$\beta^*(\phi, \epsilon) = \beta(\phi, \epsilon) \times \sigma(\phi, \epsilon) / \sigma(\phi, \epsilon = 0.5)$}.
That is: we have scaled all data for equal $\phi$ and different $\epsilon$ to the effective granular temperature that corresponds to $\epsilon = 0.5$.
The results can be observed in Fig.~\ref{fig:Tgranular_scaling} as the pink symbols, the spread among which is considerably less than the spread among the
black/gray symbols, especially for the two higher packing fractions $\phi = 0.4$ and $0.5$ (panels b and a of Fig.~\ref{fig:Tgranular_scaling}, respectively).
This confirms that $\sigma(\phi, \epsilon)$ is a good measure for an effective granular temperature.
It also supports the conceptual idea of fitting the out-of-equilibrium, steady state particle pair correlations with those of equilibrium systems,
as it is done in the \mbox{IO--ZI} methods.

\begin{figure}
\centering
\vspace{2em}
\includegraphics[width=.75\columnwidth]{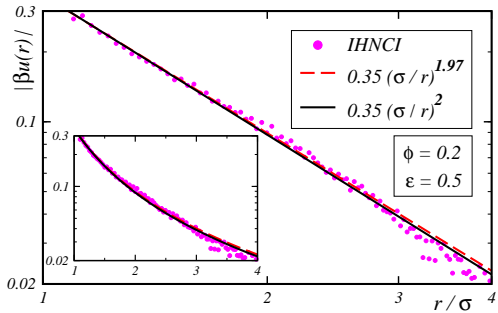}
\vspace{0em}
\caption{\label{fig:rsquare}
(Color online) Circles: Absolute value of the reduced effective IHNCI potential for $\phi = 0.2$, $\epsilon = 0.5$ (as in panel i of Fig.~\ref{fig:Main_Figure}) 
on a double logarithmic and a linear-logarithmic scale (inset).
Red dashed curve: Two-parametric fit in which both the prefactor and the exponent were allowed to vary.
Black solid curve: One-parametric fit with fixed exponent of $-2$, where only the prefactor was adjusted.   
}
\end{figure}

The observed simple shapes of $\beta u(r)$ in Fig.~\ref{fig:Main_Figure} encourage an attempt to determine the functional form of the potential,
at least for the most dilute case $\phi = 0.2$. To this end, in Fig.~\ref{fig:rsquare} we are plotting the absolute value of $\beta u(r)$ from \mbox{IHNCI},
for $\phi = 0.2$, $\epsilon = 0.5$ (as in panel i of Fig.~\ref{fig:Main_Figure}) on a double logarithmic scale and on a linear-logarithmic scale (inset of  Fig.~\ref{fig:rsquare}).
An exponential form of the potential is ruled out by the linear-logarithmic plot, where the \mbox{IHNCI} results exhibit a significant non-zero curvature.
The double logarithmic plot reveals that the \mbox{IHNCI} result is compatible with the power law $\beta u(r) = \alpha ~ r^{-2}$, with a single adjustable parameter
$\alpha$. If the exponent in the power law is allowed to vary in a non-linear regression-like fit, then an optimal exponent of $-1.97$ is obtained, providing strong
support for the hypothesized exponent of $-2$.
In default of an analytical theory for the shape of the potential, we do not want to over-interpret the results in Fig.~\ref{fig:rsquare} by stating that
the effective potential is truly of the form $\beta u(r) = \alpha ~ r^{-2}$. We merely report that our data is compatible with such a power law, and that the
theoretical justification or falsification of the power law is a rewarding task for future studies.

\section{Conclusions}\label{sec:Conclusions}

Our successful application of \mbox{IO--ZI} in its two flavors \mbox{IHNCI} and \mbox{IPYI} demonstrates that the particle correlation functions in
quasi-two-dimensional vibrated granular systems can be mapped onto those of an equivalent, truly
equilibrium system in a wide range
of granular packing fractions and restitution coefficients. The resulting effective interaction potentials exhibit a simple shape that is in line with
intuitive physical arguments. At low packing fraction, there is strong empirical evidence for the one-parametric power-law form $\beta u(r) = \alpha ~ r^{-2}$
of the effective potential. Additional analytical-theoretical work is required to support or falsify the validity of the suggested power-law form of
$\beta u(r)$. The simple \mbox{HNCI} and \mbox{PYI} methods should not be used in the determination of (effective) particle interaction potentials
as the results of these methods suffer from great systematic errors unless the particle packing fraction is very small.
Our work includes the first reported validation of \mbox{IHNCI} and \mbox{IPYI} for two-dimensional systems. Both methods are awaiting
further applications in two- and three-dimensional granular, molecular and Brownian systems.

\section*{Acknowledgements}

We acknowledge financial support from CONACyT (Grant No. 237425/2014) and PRODEP (Grant No. 511-6/17-11852).


\bibliography{PRE_Granular_Rodriguez_Heinen}

\end{document}